\documentclass[12pt]{article}
\usepackage{amsmath,amssymb,amsthm,mathrsfs,bm}
\usepackage{amscd}
\usepackage[dvipdfmx]{graphicx,color}
\usepackage{tcolorbox}
\usepackage{float}
\usepackage[all]{xy}
\usepackage{url}
\PassOptionsToPackage{linktocpage}{hyperref}
\usepackage[hyperindex,breaklinks]{hyperref}
\usepackage{titlesec}
\usepackage{caption}

\titleformat*{\section}{\normalsize\bfseries}
\titleformat*{\subsection}{\normalsize\bfseries}

\newcommand*{\affaddr}[1]{#1} 

\newcommand*{\email}[1]{\texttt{#1}}
\begin{document}
\title{\vspace{1cm}\bf\Large{\textsf {qBitcoin: A Peer-to-Peer Quantum Cash System}\\
\noindent\rule{\textwidth}{1.5pt}}}
\author{%
\bf{\textsf Kazuki Ikeda}\\
\affaddr{\small{Department of Physics, Osaka University, Japan}\\
\email{\small{kikeda@het.phys.sci.osaka-u.ac.jp, kazuki7131@gmail.com}}}\\
}
\author{%
\bf{\textsf Kazuki Ikeda}\\
\affaddr{\small{Department of Physics, Osaka University, Toyonaka, Osaka 560-0043, Japan}\\\small{\email{kikeda@het.phys.sci.osaka-u.ac.jp}}}
}
\date{}
\maketitle
\begin{abstract}
\renewcommand{\thefootnote}{\fnsymbol{footnote}}
\footnote[0]{\hspace{-6mm} To appear in Proceedings of 2018 SAI Computing Conference (SAI)}
\renewcommand{\thefootnote}{\arabic{footnote}}
A decentralized online quantum cash system, called qBitcoin, is given. We design the system which has great benefits of quantization in the following sense. Firstly, quantum teleportation technology is used for coin transaction, which prevents from the owner of the coin keeping the original coin data even after sending the coin to another. This was a main problem in a classical circuit and a blockchain was introduced to solve this issue. In qBitcoin, the double-spending problem never happens and its security is guaranteed theoretically by virtue of quantum information theory. Making a block is time consuming and the system of qBitcoin is based on a quantum chain, instead of blocks. Therefore a payment can be completed much faster than Bitcoin. Moreover we employ quantum digital signature so that it naturally inherits properties of peer-to-peer (P2P) cash system as originally proposed in Bitcoin. 
\end{abstract}
\newpage
\section{Introduction}
\renewcommand{\thefootnote}{\arabic{footnote}}
Bitcoin \cite{SN} is a decentralized online cash system and its developed versions have been widely used in the modern society as an alternative of traditional currency. However its security basically relies on cryptography based on the computation hardness assumptions in terms of classical computers and recent advances on quantum computers constitute a grave menace to the safety. It is believed that quantum computers can hasten the mining process and may crack the SHA256 hash algorithm used by Bitcoin network. Actually it is reported that the classical signature scheme used by Bitcoin is at risk and could be completely broken by quantum attack within a couple of decades \cite{2017arXiv171010377A}. Even though it may not mean Bitcoin faces immediate danger of being hacked by a quantum computer, no one will doubt there is a potential danger. Of course, the similar problem is true of banks all over the world. In order to solve this issue, we try to transplant it on a quantum network and to quantize the whole system by use of all quantum technologies which keep on evolving. 

Novelty of qBitcoin compared to Bitcoin or other quantum money schemes can be summarized as follows.
\begin{enumerate}
\item Quantum teleportation is used to transmit a coin. This prevents from double spending in a simple way without help of a blockchain. 
\item Quantum digital signature is used to verify transaction. This requires other participants to be involved in verifying signatures, hence it is compatible with P2P. 
\item Transaction can be completed much faster than Bitcoin. Each node is not a block but a point, hence transaction is immediately completed once if signature is verified.   
\end{enumerate}
We shortly explain the system in the rest of this section.  First of all, we need to consider how to transmit a coin made of quantum information. The best way, at the present, will be to employ quantum teleportation \cite{PhysRevLett.70.1895,cite-key,Furusawa706,2013Natur.500..315T}, which is succeed in transforming quantum information to remote places. Great benefit to use the quantum teleportation is that the quantum information cannot remain the original place, in other words, it is impossible for a transmitter to keep the original quantum data once if the quantum information is sent. This point is quite useful for coin transaction since any coin data should not be duplicatable, and this gives a strong advantage to qBitcoin. Quantum money we employ is not to be forged, therefore so-called the double spending problem never occurs on qBitcoin. On the other hand, this issue was crucial in Bitcoin and the blockchain system was invented to solve it. However, this causes another problem: making a block is time-consuming and one needs to wait more than 10 minutes to complete a transaction. qBitcoin is made of a network connecting points, rather than blocks. Therefore the quantized system will succeed in transmitting money much faster than Bitcoin.  

Secondly, we should mention security. As most commonly discussed, it is vital to maintain the system so that it is tolerant to hacking by a third party. In the conventional banking systems including Bitcoin, their security systems rely on cryptography whose security is guaranteed temporarily by the computational hardness assumption as represented by integer factorization algorithms and elliptic curve cryptography. However it is widely known that a quantum computer manages to break such conventional code in a short time.  So it is wise to find an alternative method. In qBitcoin, we employ quantum cryptography which is secure since it is protected by the laws of physics, hence it is secure forever in principle. Moreover owner's privacy can be perfectly protected in such a way that his/her private information is not leaked out due to blind quantum computation \cite{broadbent2009universal,cite-key2,PhysRevA.87.050301,2012NatCo...3E1036M}. This makes the system robuster by far than the conventional Bitcoin. 

Thirdly, it is also important to inherit the crucial property of Bitcoin to qBitcoin; the peer to peer (P2P) system, since this makes different from the traditional banking system, namely Bitcoin succeed in establishing an online cash system without a help of an authorized third party. We consider a novel signature verification system based on quantum digital signature. We will discuss its detail in the later section. 

This piece is orchestrated as follows. In the next section, we define the concept of a quantized coin used in qBitcoin. In section 3, we design the system of qBitcoin, in which coin transaction based on quantum teleportation and our quantum signature verification system are described in a more detailed way. In the succeeding section, we mention security based on quantum physics and some of the related works are mentioned.

\section{Coin}
A coin is defined by a pair of classical and quantum states $c_i=(r_i, |\psi_i\rangle)$, where $i$ labels a serial number and $r_i$ is a transaction record made of classical bits and transaction is done by passing $c_i$ to another person. These classical bits $r_i$ and quantum bits (qubits) $|\psi_i\rangle$ given to coins should be one-to-one correspondence so that no one can duplicate. Namely, they obey $r_i\neq r_j\Leftrightarrow  |\psi_i\rangle\neq  |\psi_j\rangle$ for all serial numbers $i,j$. What is different form the usual currency or Bitcoin is that such quantum information about the coin is hidden to anybody. The mint derivers quantum coins to those who want to trade. Owners of coins can possess the quantum state but cannot obtain full information to forge. Moreover the no-cloning theorem \cite{wootters1982single,DIEKS1982271} helps the system prevent owners from making copies of coins. Even if full quantum information is not opened to the public, one can transport the information to somebody else using quantum teleportation as explained by the succeeding section. Those serial numbers should be authenticated without contact with a central authority. As an example of such a quantum money scheme is \cite{farhi2012quantum}, which has the desired properties: 
\begin{enumerate}
\item anyone can verify money given by the mint, but cannot forge it. 
\item anyone can authenticate the serial numbers. 
\end{enumerate}

\section{Transactions}
\subsection{Remittance}
In qBitcoin, quantum information of a coin and a bank statement is transmitted by use of quantum teleportation \cite{PhysRevLett.70.1895}, which is a protocol to send quantum information to remote locations via classical information network. Experimental techniques are already established \cite{Furusawa706,2013Natur.500..315T}. The procedure to transmit a coin is as follows. Let $|\psi\rangle$ be a coin which a remitter wants to send to a receiver.   
The remitter and the receiver share a EPR pair \cite{PhysRev.47.777} and the remitter performs a Bell measurement on one of the EPR pair and $|\psi\rangle$. Then the remitter tells the outcome to the receiver via classical channel, by which the receiver can recover the information of $|\psi\rangle$ by performing a unitary operation on the other EPR pair. Through this measurement, quantum states the remitter possessed are discarded and the coin $|\psi\rangle$ is sent to the receiver. In this way, the coin data dose not remain in remitter's hand and therefore this solves the double-spending problem. For successfully communicating the information of $|\psi\rangle$, we employ a quantum key distribution (QKD) protocol and the remitter and the receiver shear a private key. The most well known is BB84 \cite{bennett1984quantum}. The remitter decode the outcome of a Bell measurement and tell it to a receiver via open channel. Decoding it, the receiver can get the coin.   
  
\begin{figure}[H]
\centering
\includegraphics[width=12cm]{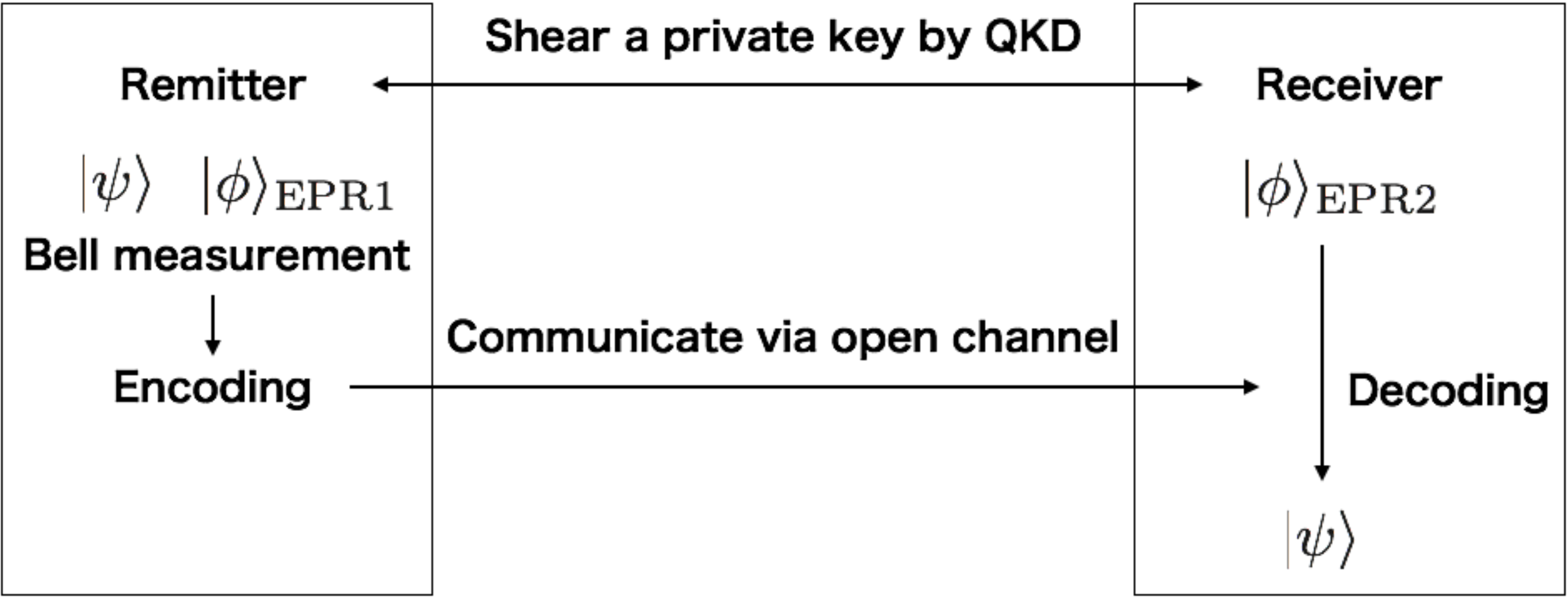}
\caption{A sketch of remitting a coin by quantum teleportation.}
\end{figure}

\subsection{Verification by use of quantum digital signature}
qBitcoin is a quantum chain equipped with a transaction system. There a coin, which is a quantum state, is to be delivered and transactions are approved once if signature of coin's owner is verified.    \if{Corresponding quantum states are assigned to owners' private keys and public keys. A private key $|\phi\rangle$ of an owner should be translated into corresponding public key $|\varphi\rangle$ in an appropriate way of quantum cryptography. Suppose the total number of owners in our market are $N$. Let $|\phi_j\rangle,|\varphi_j\rangle$ be the private key and the public key of the $j$-th owner. We normalize them as
\begin{equation}
\langle\phi_i|\phi_j\rangle=
\begin{cases}
1&i=j\\
0&i\neq j
\end{cases},~
\langle\varphi_i|\varphi_j\rangle=
\begin{cases}
1&i=j\\
0&i\neq j
\end{cases}
\end{equation}
Any private key is protected in principle thanks to the non-cloning theorem: it is impossible to make a perfect copy of an unknown pure state by an unitary operation. Generally, encoding $|\phi\rangle\to|\varphi\rangle$ is performed by a certain map from a Hilbert space to a larger Hilbert space. }\fi Security of modern digital signatures is based on the difficulty of solving a mathematical problem, such as finding the factors of large numbers (as used in the RSA algorithm). However, the task of solving these problems becomes feasible when a quantum computer is available. Moreover, traditional Bitcoin uses the classical coding, hence it is important to find an alternative way. To fix this problem, quantum digital signature schemes, which is quantum mechanical digital signature, are in development to provide protection against tampering, even from parties in possession of quantum computers and using powerful quantum cheating strategies. We employ quantum digital signature for qBitcoin. The scheme proposed by Gottesman and Chuang \cite{2001quant.ph..5032G} is the most famous. It is essentially given  by a quantum one-way function whose input $k$ is a classical bit-string and output is a corresponding quantum state $|f_k\rangle$, and inverting $k$ is impossible thanks to quantum information theory\footnote{Practically, we may also convert a classical private key $k$ into a quantum private key $|k\rangle$ and we consider a quantum one-way function $|k\rangle\mapsto|f_k\rangle$. In this way the quantum private key $|k\rangle$ is secure since nobody else but the owner can copy it by virtue of the no-cloning theorem and since it is impossible to invert $|k\rangle$ from $|f_k\rangle$.}:
\begin{align}\notag
\begin{aligned}
k&\mapsto |f_k\rangle~~\text{\bf easy}\\
|f_k&\rangle\mapsto k~~\text{\bf impossible}
\end{aligned}
\end{align} 
There is a key distribution method which secures the quantum digital signature, which allows us to design qBitcoin on this scheme. Unlike Bitcoin, there is a limitation of issuing public keys of qBitcoin owners. If $k$ has length $L=O(2^n)$, then one should make $T$ copies of $|f_k\rangle$ so that $L-nT\gg 1$ since if one observes too many copies of the public key, then a chance of successfully guessing the initial private key $k$ becomes big, which is a consequence of Holevo's theorem \cite{HOLEVO1977273}. Therefore we should limit the number of copies of any public key for the map $k\mapsto|f_k\rangle$ being a quantum one-way function, and such public keys are distributed to the corresponding number of other participants so that they can verify the signature.  
 
The system of the quantum digital signature requires other participants to be involved in verifying the signature, hence it is naturally compatible with the concept of Bitcoin as a P2P electric cash system. Moreover hash algorithm is not be used and a transmitter must sign every single bit of a invoice. The procedure of a transaction is illustrated and described as follows. 
\begin{figure}[H]
\centering
\includegraphics[width=10cm]{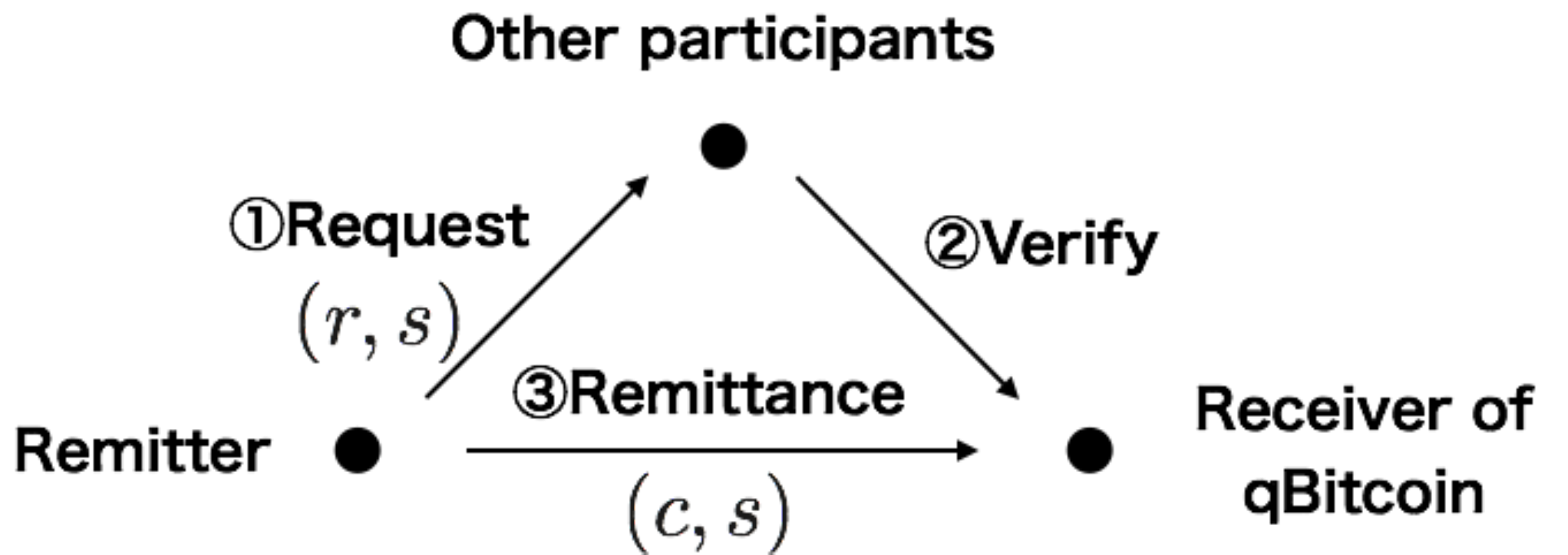}
\end{figure}
\begin{enumerate}
\item A remitter and a receiver shear a private key using a QKD protocol, such as BB84. 
\item Remittance is accepted if and only if a coin $c=(r, |\psi\rangle)$ has the same serial numbers with respect to $r$ and $|\psi\rangle$. 
\item The remitter encodes $|\psi\rangle$.  
\item The remitter sends a remittance request which includes the record $r$, the signature $s=(k, |f_k\rangle)$, and the encoded information of $|\psi\rangle$ to randomly chosen participants who are able to receive the corresponding public key $|f_k\rangle$.   
\item The receivers of the request verify the signature.
\item If the signature is approved, then one of the receivers renews the record and open it to public. (Some reward will be given to the one(s) for incentive. )
\item The receiver of qBitcoin can receive the coin by decoding the information of $c$.  
\end{enumerate}

\if{
\textcolor{blue}{Alice chooses a number of pairs of $L$-bit strings $\{k_0^i,k_1^i\},~1\le i\le M$. The $k_0=(k_0^1,\cdots,k_0^M)$ are used to sign a message $b = 0$, and the $k_1=(k_1^1,\cdots,k_1^M)$ are used to sign $b = 1$. Note $k_0^i$ and $k_1^i$ are chosen independently and randomly for each $i$, and $M$ keys are used to sign each bit. $M$ is the security parameter; the protocol is exponentially secure in $M$ when the other parameters are fixed. The states $\{|f_{k_0^i}\rangle, |f_{k_1^i}\rangle\}$ (for each $i$) will then be Alice’s public keys for an appropriate quantum one-way function $f$. It is important to note that the creation of these keys is up to Alice, since unknown quantum states cannot be perfectly copied, according to the no-cloning theorem. We begin by making the simplifying assumption that all recipients have received correct and identical copies of Alice’s public keys; we will revisit this assumption later in the paper.
All participants in the protocol will know how to implement the map $k \mapsto |f_k \rangle$. All participants will also know two numbers, $c_1$ and $c_2$, thresholds for acceptance and rejection used in the protocol. A bound on the allowed value of $c_2$ will be given as part of the proof of security, below. $c_1$ can be zero in the absence of noise; the gap $c_2-c_1$ limits Alice’s chance of cheating. We assume perfect devices and channels throughout this paper, but our protocol still works in the presence of weak noise by letting $c_1$ be greater than zero, and with other minor adjustments. We further require that Alice limits distribution of the public keys so that $T < L/n$ copies of each key are available (recall that $|f_k\rangle$ is an $n$ qubit state).
Alice can now send a single-bit message $b$ using the following procedure:
\begin{enumerate}
\item Alice sends the signed message$(b,k_b^1,k_b^2,...,k_b^M)$ over an insecure classical channel. Thus, Alice reveals the identity of half of her public keys.
\item Each recipient of the signed message checks each of the revealed public keys to verify that $kbi  \mapsto |f_{k_b^i}\rangle$. Recip- ient j counts the number of incorrect keys; let this be sj.
\item Recipient j accepts the message as valid and transfer-
able (result 1-ACC) if $s_j \le c_1 M$, and rejects it as invalid (result REJ) if $s_j \ge c_2M$. If $c_1M < s_j < c_2M$, recipient j concludes the message is valid but not necessarily transferable to other recipients (result 0-ACC).
\item Discard all used and unused keys.
\end{enumerate}
When $s_j$ is large, the message has been heavily tampered with, and may be invalid. When it is small, the message cannot have been changed very much from what Alice sent. $s_j$ is similar for all recipients, but need not be identical. As we shall see below, the thresholds $c_1$ and $c_2$ separate values of $s_j$ into different domains of security. Forgery is prevented by $c_2$, and cheating by Alice is prevented by a gap between $c_2$ and $c_1$.}}\fi

\if{
\section{Much faster payment verification}  
In comparison with the current Bitcoin system, here we explain that qBitcoin makes payment much faster. The reason why it takes more than ten minutes to approve payment on Bitcoin is because mining coins takes time. It is almost impossible to shorter this procedure unless redesigning the system since the system is based on the classical computational ways. Meanwhile it is possible for qBitcoin owners to making up accounts within 0.01 seconds thanks to a quantum algorithm of qBitcoin. 

In Bitcoin, mining coins is done by the proof-of-work, in which computers of participants are required to find a randomly assigned vale (32 bites, for example). Therefore in order to make payment faster, it is essential to find an algorithm which enables to search a random vale as first as possible. In the conventional ways of Bitcoin, users usually try to find a correct value by increasing initial numbers one by one like they find unknown combination numbers of a chain lock. This is the reason mining coins takes time. 

We consider this problem in a general viewpoint. Let $x$ be a $n$-bites value $x\in \{0,1\}^n$ and $x_0$ be the unique value we are looking for. We define the function $f:\{0,1\}^n\to \{0,1\}$ by
\begin{equation} 
f(x)=
\begin{cases}
1& x=x_0\\
0&x\neq x_0
\end{cases}
\end{equation}
There are $N=2^n$ different combinations of numbers and it will take $N-1$ steps in the worst case, if we repeat the classical way, namely we start from $00\cdots0$ (or $|00\cdots0\rangle$ as a quantum state) and increase numbers one by one until we find $x_0$. 

An excellent quantum algorithm for this problem is given \cite{Grover:1996rk}, which allows us to adopt the proof-of-work even for qBitcoin initiatively. Grover's algorithm guarantees to search $x_0$ within about $\sqrt{N}$ steps with a probability of $1-1/N$. Hence in order to find $x_0$ with probability of almost 100\% for large $N$, the quantum algorithm makes a payment $\frac{N-1}{\sqrt{N}}$ times faster than the classical one. Therefore if $n=32$, one will take roughly $0.009$ seconds for a payment. 
}\fi

\section{Security and Privacy} 
Here we mention the security of qBitcoin against several scenarios of cheaters. We first assume a remitter is a cheater. The case where the remitter copies a coin data and sends them more than two persons is already rejected since the remitter dose not know the corresponding quantum state of a coin and so any copy of a coin cannot be generated. Moreover no third party can steal a coin by virtue of security of a QKD protocol. Namely, for trying to steal a coin, a third party needs to wiretap communication between a remitter and a receiver and to get a private key sheared by them, however, this attempt never becomes successful in principle since a QKD should be protected by laws of physics.    

qBitcoin strengths owner's privacy better than Bitcoin. As mentioned in the original paper, the risk of Bitcoin is that if the owner of a key is revealed, linking could reveal other transactions that belonged to the same owner. qBitcoin solves this point by using quantum digital signature of type proposed by Gottesman and Chuang. There owners' private keys are not secret and any key is not link to themselves. Moreover owners can generate keys as many as they like hence it is impossible even to guess to whom they belong.   

In practice, it will be also important to consider how one who does not oppose a quantum computer can use quantum money since only a limited number of people will be able to have a first generation quantum computer. In this case, one who wants to trade quantum money will be forced to access to a quantum computer at an exchange. Here question is how to protect owner's privacy. Fortunately we have a good solution to this issue. Namely, so called blind quantum computation \cite{broadbent2009universal,cite-key2} is a secure protocol which enables a client (Alice) without a quantum computer to delegate her quantum computation to a server (Bob) with a fully fledged quantum technology in such a way that Bob cannot obtain any information about Alice's actual input, output, and algorithm. This protocol is very secure and robust against realistic noise thanks to topological blind quantum computation \cite{PhysRevA.87.050301,2012NatCo...3E1036M}. Applying this protocol to qBitcoin, owner's privacy should be perfectly protected. On the other hand, establishing classical blind computation in a practical manner is still open problem. That is to say, in the conventional Bitcoin, private information of coin owners is stored at market place and can be leaked out. 

\section{Related works}
The attempt to making a money system based on quantum mechanics has a long history. It is believed that Wiesner made a prototype in about 1970 (published in 1983) \cite{Wiesner:1983:CC:1008908.1008920}, in which quantum money that can be verified by a bank is given. In his scheme, quantum money was secure in the sense that it cannot be copied due to the no-cloning theorem, however there were several problems. For example a bank need to maintain a giant data base to store a classical information of quantum money.  
Aaronson proposed a quantum money scheme where public key was used to verify a banknote \cite{2011arXiv1110.5353A} and later his scheme was developed in \cite{2012arXiv1203.4740A}. There is a survey on trying to quantize Bitcoin \cite{2016arXiv160401383J} based on a classical blockchain system and a classical digital signature protocol proposed in \cite{2012arXiv1203.4740A}. However, all of those works rely on classical digital signature protocols and classical coin transmission system, hence computational hardness assumptions are vital to their systems. In other words, if a computer equipped with ultimate computational ability appears someday, the money systems above are in danger of collapsing, as the bank systems today face.  

\section{Conclusion and Future Work}
What we presented in this article can be summarized as follows. qBitcoin is a decentralized online quantum cash system and the significant change is that quantum states as coins are exchanged in addition to transaction descriptions. This secures traditional Bitcoin much better than previous good works in the following sense that no one can cheat transactions and no third party can steal any information illegally according to the laws of physics. Moreover, privacy of owners are perfectly protected. By virtue of the successful blind quantum computation, we also emphasize that one without full quantum technology can trade quantum money on qBitcoin without having to worry about any privacy problem. This is never achievable on Bitcoin. Furthermore qBitcoin can complete a transaction faster than Bitcoin since a blockchain is replaced by a quantum chain.

Regarding future work, it will be interesting to invent a quantum blockchain, which accommodates quantum information and remittance requests are accepted by rigorous peer review. Straightforward implementation of a blockchain with quantum states is complicated by the fact the original protocol is based on sending several copies of received messages to other participants. For instance, Alice sends Charlotte and David a message that she received from Bob. It is a nontrivial task due to the constraints of the no-cloning theorem.
\section*{Acknowledgement}
I was benefited by discussions with Keisuke Fujii, Hayato Hirai, Hiroto Hosoda and Teppei Nakano. A part of this work was completed while I stayed at Sydney for ICML 2017. I thank the organisers and I am most grateful to the Tjandras, my host family, for their constant hospitality.   
\bibliographystyle{utphys}
\bibliography{Ref}

\providecommand{\href}[2]{#2}\begingroup\raggedright\begin{thebibliography}{10}

\bibitem{SN}
S.~Nakamoto, ``Bitcoin: A peer-to-peer electric cash system,''.
  \url{http://www.bitcoin.org/bitcoin.pdf}.

\bibitem{2017arXiv171010377A}
D.~{Aggarwal}, G.~K. {Brennen}, T.~{Lee}, M.~{Santha}, and M.~{Tomamichel},
  ``{Quantum attacks on Bitcoin, and how to protect against them},'' {\em ArXiv
  e-prints} (Oct., 2017) , \href{http://arxiv.org/abs/1710.10377}{{\ttfamily
  arXiv:1710.10377 [quant-ph]}}.

\bibitem{PhysRevLett.70.1895}
C.~H. Bennett, G.~Brassard, C.~Cr\'epeau, R.~Jozsa, A.~Peres, and W.~K.
  Wootters, ``Teleporting an unknown quantum state via dual classical and
  einstein-podolsky-rosen channels,''
  \href{http://dx.doi.org/10.1103/PhysRevLett.70.1895}{{\em Phys. Rev. Lett.}
  {\bfseries 70} (Mar, 1993) 1895--1899}.
  \url{https://link.aps.org/doi/10.1103/PhysRevLett.70.1895}.

\bibitem{cite-key}
D.~Bouwmeester, J.-W. Pan, K.~Mattle, M.~Eibl, H.~Weinfurter, and A.~Zeilinger,
  ``Experimental quantum teleportation,'' {\em Nature} {\bfseries 390}
  no.~6660, (12, 1997) 575--579.

\bibitem{Furusawa706}
A.~Furusawa, J.~L. S{\o}rensen, S.~L. Braunstein, C.~A. Fuchs, H.~J. Kimble,
  and E.~S. Polzik, ``Unconditional quantum teleportation,''
  \href{http://dx.doi.org/10.1126/science.282.5389.706}{{\em Science}
  {\bfseries 282} no.~5389, (1998) 706--709},
  \href{http://arxiv.org/abs/http://science.sciencemag.org/content/282/5389/706.full.pdf}{{\ttfamily
  http://science.sciencemag.org/content/282/5389/706.full.pdf}}.
  \url{http://science.sciencemag.org/content/282/5389/706}.

\bibitem{2013Natur.500..315T}
S.~Takeda, T.~Mizuta, M.~Fuwa, P.~van Loock, and A.~Furusawa, ``Deterministic
  quantum teleportation of photonic quantum bits by a hybrid technique,'' {\em
  Nature} {\bfseries 500} no.~7462, (08, 2013) 315--318,
  \href{http://arxiv.org/abs/1402.4895}{{\ttfamily arXiv:1402.4895
  [quant-ph]}}. \url{http://dx.doi.org/10.1038/nature12366}.

\bibitem{broadbent2009universal}
A.~Broadbent, J.~Fitzsimons, and E.~Kashefi, ``Universal blind quantum
  computation,'' in {\em Foundations of Computer Science, 2009. FOCS'09. 50th
  Annual IEEE Symposium on}, pp.~517--526, IEEE.
\newblock 2009.

\bibitem{cite-key2}
S.~Barz, J.~F. Fitzsimons, E.~Kashefi, and P.~Walther, ``Experimental
  verification of quantum computation,'' {\em Nat Phys} {\bfseries 9} no.~11,
  (11, 2013) 727--731.

\bibitem{PhysRevA.87.050301}
T.~Morimae and K.~Fujii, ``Blind quantum computation protocol in which alice
  only makes measurements,''
  \href{http://dx.doi.org/10.1103/PhysRevA.87.050301}{{\em Phys. Rev. A}
  {\bfseries 87} (May, 2013) 050301}.
  \url{https://link.aps.org/doi/10.1103/PhysRevA.87.050301}.

\bibitem{2012NatCo...3E1036M}
T.~{Morimae} and K.~{Fujii}, ``{Blind topological measurement-based quantum
  computation},'' \href{http://dx.doi.org/10.1038/ncomms2043}{{\em Nature
  Communications} {\bfseries 3} (Sept., 2012) 1036},
  \href{http://arxiv.org/abs/1110.5460}{{\ttfamily arXiv:1110.5460
  [quant-ph]}}.

\bibitem{wootters1982single}
W.~K. Wootters and W.~H. Zurek, ``A single quantum cannot be cloned,'' {\em
  Nature} {\bfseries 299} no.~5886, (1982) 802--803.

\bibitem{DIEKS1982271}
D.~Dieks, ``Communication by epr devices,''
  \href{http://dx.doi.org/http://dx.doi.org/10.1016/0375-9601(82)90084-6}{{\em
  Physics Letters A} {\bfseries 92} no.~6, (1982) 271 -- 272}.
  \url{http://www.sciencedirect.com/science/article/pii/0375960182900846}.

\bibitem{farhi2012quantum}
E.~Farhi, D.~Gosset, A.~Hassidim, A.~Lutomirski, and P.~Shor, ``Quantum money
  from knots,'' in {\em Proceedings of the 3rd Innovations in Theoretical
  Computer Science Conference}, pp.~276--289, ACM.
\newblock 2012.

\bibitem{PhysRev.47.777}
A.~Einstein, B.~Podolsky, and N.~Rosen, ``Can quantum-mechanical description of
  physical reality be considered complete?,''
  \href{http://dx.doi.org/10.1103/PhysRev.47.777}{{\em Phys. Rev.} {\bfseries
  47} (May, 1935) 777--780}.
  \url{https://link.aps.org/doi/10.1103/PhysRev.47.777}.

\bibitem{bennett1984quantum}
C.~H. Bennett and G.~Brassard, ``Quantum cryptography: Public key distribution
  and con tos5,''.

\bibitem{2001quant.ph..5032G}
D.~{Gottesman} and I.~{Chuang}, ``{Quantum Digital Signatures},'' {\em eprint
  arXiv:quant-ph/0105032} (May, 2001) ,
  \href{http://arxiv.org/abs/quant-ph/0105032}{{\ttfamily quant-ph/0105032}}.

\bibitem{HOLEVO1977273}
A.~Holevo, ``Problems in the mathematical theory of quantum communication
  channels,''
  \href{http://dx.doi.org/http://dx.doi.org/10.1016/0034-4877(77)90010-6}{{\em
  Reports on Mathematical Physics} {\bfseries 12} no.~2, (1977) 273 -- 278}.
  \url{http://www.sciencedirect.com/science/article/pii/0034487777900106}.

\bibitem{Wiesner:1983:CC:1008908.1008920}
S.~Wiesner, ``Conjugate coding,''
  \href{http://dx.doi.org/10.1145/1008908.1008920}{{\em SIGACT News} {\bfseries
  15} no.~1, (Jan., 1983) 78--88}.
  \url{http://doi.acm.org/10.1145/1008908.1008920}.

\bibitem{2011arXiv1110.5353A}
S.~{Aaronson}, ``{Quantum Copy-Protection and Quantum Money},'' {\em ArXiv
  e-prints} (Oct., 2011) , \href{http://arxiv.org/abs/1110.5353}{{\ttfamily
  arXiv:1110.5353 [quant-ph]}}.

\bibitem{2012arXiv1203.4740A}
S.~{Aaronson} and P.~{Christiano}, ``{Quantum Money from Hidden Subspaces},''
  {\em ArXiv e-prints} (Mar., 2012) ,
  \href{http://arxiv.org/abs/1203.4740}{{\ttfamily arXiv:1203.4740
  [quant-ph]}}.

\bibitem{2016arXiv160401383J}
J.~{Jogenfors}, ``{Quantum Bitcoin: An Anonymous and Distributed Currency
  Secured by the No-Cloning Theorem of Quantum Mechanics},'' {\em ArXiv
  e-prints} (Apr., 2016) , \href{http://arxiv.org/abs/1604.01383}{{\ttfamily
  arXiv:1604.01383 [quant-ph]}}.

\end{thebibliography}\endgroup
\end{document}